\documentclass[12pt]{article}
\usepackage[english]{babel}
\usepackage[dvips]{epsfig}
\usepackage{cite}
\usepackage{amsmath,amssymb}
\usepackage{epsfig}

\newcommand{\ep}{\varepsilon}

\begin{document}
\thispagestyle{empty}
\centerline{DESY 13--071\hfill ISSN 0418--9833}
\centerline{May 2013\hfill}
\vspace*{2.0cm}
\begin{center}
 {\large \bf
{\sf HYPERDIRE} \\
HYPERgeometric functions DIfferential REduction: \\
MATHEMATICA based packages for
differential reduction of generalized hypergeometric functions \\
${}_pF_{p-1}, F_1,F_2,F_3,F_4$
 }
\end{center}
 \vspace*{0.8cm}

\begin{center}
{\sc Vladimir~V.~Bytev$^{a,b}$}, \quad
{\sc Mikhail~Yu.~Kalmykov$^{a,b}$}, \quad
{\sc Bernd~A.~Kniehl$^{a}$ } \\
 \vspace*{1.0cm}
{\normalsize $^{a}$ II. Institut f\"ur Theoretische Physik, Universit\"at Hamburg,}\\
{\normalsize Luruper Chaussee 149, 22761 Hamburg, Germany} \\
{\normalsize $^{b}$ Joint Institute for Nuclear Research,} \\
{\normalsize $141980$ Dubna (Moscow Region), Russia}
\end{center}

\begin{abstract}
{\bf HYPERDIRE} is a project devoted to the creation of a set of Mathematica based programs
for the differential reduction of hypergeometric functions.
The current version includes two parts:
one, {\bf pfq}, is relevant for manipulations of hypergeometric functions $_{p+1}F_{p}$,
and the second one, {\bf AppellF1F4}, for manipulations with Appell hypergeometric functions
$F_1,F_2,F_3,F_4$
of two variables.

\medskip

\noindent
PACS numbers: 02.30.Gp, 02.30.Lt, 11.15.Bt, 12.38.Bx\\
Keywords: Hypergeometric functions; Differential reduction

\end{abstract}

\newpage

{\bf\large PROGRAM SUMMARY}
\vspace{4mm}
\begin{sloppypar}
\noindent   {\em Title of program\/}: {\sf HYPERDIRE} \\[2mm]
   {\em Version\/}: 1.0.0
   {\em Release\/}: 1.0.0
   {\em Catalogue number\/}: \\[2mm]
   {\em Program obtained from\/ {\tt https://sites.google.com/site/loopcalculations/home}}:
   {\tt } \\[2mm]
   {\em E-mail: bvv@jinr.ru} \\[2mm]
   {\em Licensing terms\/}: GNU General Public Licence  \\[2mm]
   {\em Computers\/}: all computers running Mathematica \\[2mm]
   {\em Operating systems\/}:  operating systems running Mathematica\\[2mm]
   {\em Programming language\/}: {\tt Mathematica  } \\[2mm]
   {\em Keywords\/}:  Generalized Hypergeometric functions, Appell functions, Feynman integrals. \\[2mm]
   {\em Nature of the problem\/}:
                  Reduction of hypergeometric functions $_{p}F_{p-1}, F_1, F_2, F_3, F_4$ to sets of basis functions.
                \\[2mm]
   {\em Method of solution\/}: Differential reduction \\[2mm]
   {\em Restriction on the complexity of the problem}: none \\[2mm]
   {\em Typical running time}:  Depending on the complexity of problem.
\end{sloppypar}
%
\newpage

{\bf\large LONG WRITE-UP}
\vspace{4mm}

\section{Introduction}
Multiple hypergeometric functions \cite{Gauss,appell,bateman, Gelfand}
play an important role in many branches of science.
In particular, a large class of Feynman diagrams are expressed in terms
of Horn-type hypergeometric functions \cite{FD}.

Let us consider a multiple series:
\begin{eqnarray}
H(\vec{\gamma};\vec{\sigma};\vec{x})
=
\sum_{m_1,m_2,\cdots, m_r=0}^\infty
\Biggl(
\frac{
\Pi_{j=1}^K
\Gamma\left(\sum_{a=1}^r \mu_{ja}m_a+\gamma_j \right)
\Gamma^{-1}(\gamma_j)
}
{
\Pi_{k=1}^L
\Gamma\left( \sum_{b=1}^r \nu_{kb}m_b+\sigma_k \right)
\Gamma^{-1}(\sigma_k)
}
\Biggr)
x_1^{m_1} \cdots x_r^{m_r} \;,
\label{H}
\end{eqnarray}
with
$
\mu_{ab}, \nu_{ab} \in \mathbb{Z},\
\gamma_j,\sigma_k \in \mathbb{C}.
$
The sequences $\vec{\gamma}=(\gamma_1,\cdots, \gamma_K)$
and $\vec{\sigma}=(\sigma_1,\cdots, \sigma_L)$ are called {\it upper }
and {\it lower} parameters of the hypergeometric function, respectively.

Let
$
\vec{e}_j = (0,\cdots,0,1,0,\cdots,0)
$
denote the unit vector with unity in its $j^{\rm th}$ entry, and let us define
$
\vec{x}^{\vec{m}} = x_1^{m_1} \cdots x_r^{m_r}
$
for any integer multi-index
$\vec{m} = (m_1, \cdots, m_r)$.
Two functions of type (\ref{H})
with sets of parameters shifted by unity,
$H(\vec{\gamma}+\vec{e_c};\vec{\sigma};\vec{x})$ and
$H(\vec{\gamma};\vec{\sigma};\vec{x})$,
are related by a linear differential operator:
\begin{eqnarray}
H(\vec{\gamma}+\vec{e_c};\vec{\sigma};\vec{x})
& = &
\frac{1}{\gamma_c}
\left ( \sum_{a=1}^r \mu_{ca} x_a \frac{\partial}{\partial x_a}+\gamma_c \right)
H(\vec{\gamma};\vec{\sigma};\vec{x})
\equiv
U_{[\gamma_c \to \gamma_c+1]}^+
H(\vec{\gamma},\vec{\sigma}, \vec{x})
\;.
\label{do1}
\end{eqnarray}
Similar relations also exist for the lower parameters:
\begin{eqnarray}
H(\vec{\gamma};\vec{\sigma}-\vec{e}_c;\vec{x})
& = &
\frac{1}{\sigma_c \!-\! 1}
\left(
\sum_{b=1}^r
\nu_{cb} x_b \frac{\partial}{\partial x_b} \!+\! \sigma_c \!-\! 1
\right)
H(\vec{\gamma};\vec{\sigma};\vec{x})
\equiv
L_{[\sigma_c \to \sigma_c-1]}^-
H(\vec{\gamma};\vec{\sigma};\vec{x})
\;.
\label{do2}
\end{eqnarray}
The linear differential operators $U_{\gamma_c \to \gamma_c+1}^+$, $L_{\sigma_c \to \sigma_c-1}^-$ are
called the  {\it step-up} and {\it step-down} operators for the
upper and lower indices, respectively.
If additional step-down and step-up operators
$U_{\gamma_c}^-$, $L_{\sigma_c}^+$ satisfying
$$
U_{[\gamma_{c}+1 \to \gamma_c ]}^- U_{[\gamma_c \to \gamma_c+1]}^+ H(\vec{\gamma},\vec{\sigma},\vec{x}) =
L_{[\sigma_{c}-1 \to \sigma_{c}]}^+ L_{[\sigma_c \to \sigma_c-1]}^- H(\vec{\gamma},\vec{\sigma},\vec{x}) =
H (\vec{\gamma},\vec{\sigma},\vec{x})
$$
({\it i.e.}, the inverses of  $U_{\gamma_c}^+$, $L_{\sigma_c}^-$)
are constructed, we can combine these operators to shift the parameters of the
hypergeometric function by any integer.
This process of applying $U_{\gamma_c}^\pm, L_{\sigma_c}^\pm$ to shift the
parameters by integers is called
{\bf differential reduction} of a hypergeometric function.

In this way, the Horn-type structure provides an opportunity to reduce
hypergeometric functions to a set of basis functions with
parameters differing from the original values by integer shifts:
\begin{equation}
P_0(\vec{x})
H(\vec{\gamma}+\vec{k};\vec{\sigma}+\vec{l};\vec{x})
=
\sum_{m_1, \cdots, m_p=0}^{\sum{|k_i|+\sum|l_i|}} P_{m_1, \cdots, m_r} (\vec{x})
\left( \frac{\partial}{\partial x_1} \right)^{m_1} \cdots
\left( \frac{\partial}{\partial x_r} \right)^{m_r}
H(\vec{\gamma};\vec{\sigma};\vec{x}) \;,
\label{reduction}
\end{equation}
where $P_0(\vec{x})$ and $P_{m_1, \cdots, m_p}(\vec{x})$ are polynomials with
respect to $\vec{\gamma},\vec{\sigma},$ and $\vec{x}$, and $\vec{k},\vec{l}$
are lists of integers.

Algebraic relations between the functions
$H(\vec{\gamma},\vec{\sigma};\vec{x})$
with parameters shifted by integers are called {\bf contiguous relations}.
The development of systematic techniques for the solution of
contiguous relations has a long history.
It was started by Gauss, who described the reduction
for the $_2F_1$ hypergeometric function in 1823 \cite{Gauss}.
Numerous papers have since been published \cite{contiguous,mullen,singal}
on this problem.
An algorithmic solution was found by Takayama in Ref.\ \cite{theorem},
and those methods have been extended later in a series of 
publications \cite{takayama,Japan} (see also Refs.~\cite{Ore,HYP,creative}).

Let us recall that any hypergeometric function
can be considered to be the solution of a proper system of partial
differential equations (PDEs).  In particular, for a Horn-type hypergeometric
function, the system of PDEs can be derived from the coefficients of the series 
$$
H = \sum_{\vec{m}} C(\vec{m}) \vec{x}^{\vec{m}}.
$$
In this case, the ratio of two coefficients can be represented as a ratio of
two polynomials,
\begin{equation}
\frac{C(\vec{m}+e_j)}{C(\vec{m})}  =  \frac{P_j(\vec{m})}{Q_j(\vec{m})} 
= 
\Pi_{j=1}^K
\frac{
\Gamma\left(\sum_{a=1}^r \mu_{ja}m_a \!+\! \mu_{ja} \delta_{ai} \!+\! \gamma_j \right)
}
{
\Gamma\left(\sum_{a=1}^r \mu_{ja}m_a \!+\! \gamma_j \right)
}
\Pi_{k=1}^L
\frac
{
\Gamma\left( \sum_{b=1}^r \nu_{kb}m_b \!+\! \sigma_k \right)
}
{
\Gamma\left( \sum_{b=1}^r \nu_{kb}m_b \!+\! \nu_{kb} \delta_{bi} \!+\! \sigma_k \right)
}
\;,
\label{pre-diff}
\end{equation}
so that the Horn-type hypergeometric function satisfies the following system
of differential equations:
\begin{equation}
0 =
D_j (\vec{\gamma},\vec{\sigma},\vec{x})
H(\vec{\gamma},\vec{\sigma}, \vec{x})
=
\left[
Q_j\left(
\sum_{k=1}^r x_k\frac{\partial}{\partial x_k}
\right)
\frac{1}{x_j}
-
 P_j\left(
\sum_{k=1}^r x_k\frac{\partial}{\partial x_k}
\right)
\right]
H(\vec{\gamma},\vec{\sigma}, \vec{x}) \;,
\label{diff}
\end{equation}
where $j=1, \ldots, r$.
It was pointed out in several publications \cite{reduction:our:1,reduction:our:2,expansion:our} that
(i) the differential reduction algorithm, Eq.~(\ref{reduction}),
can be applied to the reduction of Feynman diagrams to some subsets of basis
hypergeometric functions with well-known analytical properties \cite{reduction:our:1,reduction:our:2};
(ii) the system of differential equations, Eq.~(\ref{diff}), can be also used for the construction of
so-called $\varepsilon$ expansions of hypergeometric functions about rational values of parameters
via the direct solution of the systems of differential equations \cite{expansion:our}.
This is another motivation for creating a package for the manipulation of the parameters of
Horn-type hypergeometric functions.

The aim of this paper is to present the {\it Mathematica} \cite{math}
based package {\bf HYPERDIRE} for the differential reduction of the Horn-type hypergeometric
function with arbitrary values of parameters to a set of basis functions.
The current version consists of two parts:
one, {\bf pfq}, for the manipulation of hypergeometric functions, $_{p+1}F_{p}$,
and the second one, {\bf AppellF1F4}, for the manipulation
of Appell functions, $F_1,F_2,F_3,F_4$.
The algorithm of differential reduction for other functions can be implemented
as an additive module.

In contrast to the recent programs written by members of computational particles physics
community \cite{recent,lintz,expansion:huber-maitre,NumExp,sumino,greynat},
the aim of our package  is the manipulation of hypergeometric functions without the construction
of $\varepsilon$ expansions \cite{nested,our:expansion}.

The preliminary version of ${\bf pfq}$ was presented in Ref.~\cite{our:hep} and is available in Ref.~\cite{MKL:hyper}.
The latest version is available in Ref.~\cite{bytev:hyper}.
%
%
%
%
\section{Differential-reduction algorithm for generalized hypergeometric function ${}_{p+1}F_{p}$}
\subsection{General consideration}
\label{definition}
Let us consider the generalized hypergeometric function, $_pF_q(a;b;z)$, defined around $z=0$ by a series
\begin{equation}
{}_pF_q(\vec{a};\vec{b};z)
\equiv
{}_{p}F_q \left( \begin{array}{c|}
\vec{a} \\
\vec{b}
\end{array}~ z \right)
= \sum_{k=0}^\infty \frac{z^k}{k!} \frac{\Pi_{i=1}^p (a_i)_k}{\Pi_{j=1}^q (b_j)_k} \;,
\label{hypergeometric}
\end{equation}
where
$(a)_k$ is a Pochhammer symbol, $(a)_k = \Gamma(a+k)/\Gamma(a)$.
The sequences $\vec{a}=(a_1,\cdots, a_p)$
and $\vec{b}=(b_1,\cdots, b_q)$ are called the upper and lower parameters of hypergeometric functions, respectively.
In terms of  the operator $\theta$:
\begin{equation}
\theta = z \frac{d}{d z} \;,
\label{theta}
\end{equation}
the differential equation for the hypergeometric function ${}_pF_{q}$ can be written as
\begin{equation}
\left[
z \Pi_{i=1}^p (\theta + a_i)
- \theta \Pi_{i=1}^q (\theta + b_i-1)
\right]
{}_pF_q(\vec{a}; \vec{b}; z) = 0.
\label{diffpFq}
\end{equation}

\subsection{Differential reduction}
The differential reduction for these functions 
was analyzed in details in Ref.~\cite{reduction:our:2}.
Here we recall some of the main relations relevant to our program.

The universal differential operators, Eqs.~(\ref{do1}) and (\ref{do2}),
have the following form:
\begin{eqnarray}
{}_pF_q(a_1+1, \vec{a};\vec{b};z) & = &
B_{a_1}^{+} {}_pF_q(a_1, \vec{a};\vec{b};z)
 =
\frac{1}{a_1}
\left(\theta \!+\! a_1 \right)
{}_pF_q(a_1, \vec{a};\vec{b};z) \;,
\label{universal:a}
\\
{}_pF_q(\vec{a};b_1-1,\vec{b};z) & = &
H_{b_1}^{-} {}_pF_q(\vec{a}; b_1,\vec{b};z)
=
\frac{1}{b_1 \!-\! 1}
\left(\theta  \!+\! b_1 \!-\! 1 \right)
{}_pF_q(\vec{a};b_1,\vec{b};z) \;,
\label{universal:b}
\end{eqnarray}
where the operators $B_{a_1}^{+} (H_{b_1}^{-})$ are called  the step-up (step-down) operators
for the upper (lower) parameters of hypergeometric functions.
This type of operators were explicitly constructed for the hypergeometric
function ${}_{p+1}F_{p}$ by Takayama in Ref.~\cite{takayama}.
For completeness, we reproduce his result here:
\begin{eqnarray}
{}_{p+1}F_p(a_i-1, \vec{a}; \vec{b}; z)  & = & B_{a_i}^{-} ~{}_{p+1}F_p(a_i, \vec{a}; \vec{b}; z) \;,
\nonumber \\
{}_{p+1}F_p(\vec{a}; b_i+1, \vec{b}; z)  & = & H_{b_i}^{+} ~{}_{p+1}F_p(\vec{a}; b_1, \vec{b}; z) \;,
\end{eqnarray}
where
\begin{eqnarray}
B_{a_i}^{-} & = & - \frac{a_i}{c_i}
\left.
\left[
t_i (\theta) - z \Pi_{j\neq i} (\theta + a_j )
\right]
\right|_{a_i \to a_i - 1} \;,
\nonumber \\
c_i & = & -a_i \Pi_{j=1}^{p} (b_j-1-a_i) \; ,
\nonumber \\
t_i (x)  & = & \frac{x \Pi_{j=1}^p (x + b_j - 1) - c_i }{x+a_i}
=
\sum_{j=0}^p P^{(p)}_{p-j}(\{b_r\!-\!1\}) \frac{\left[ x^{j+1} \!-\! (-a_i)^{j+1} \right]}{x\!+\!a_i}
\nonumber \\
& =   &
\sum_{j=0}^p P^{(p)}_{p-j}(\{b_r\!-\!1\}) \sum_{k=0}^j x^{j-k}(-a_i)^k \;,
\label{diff:oper:1}
\end{eqnarray}
\begin{eqnarray}
H_{a_i}^{+} & = & \frac{b_i-1}{d_i}
\left.
\left[
\frac{d}{dz} \Pi_{j\neq i} (\theta + b_j - 1 )
- s_i (\theta)
\right]
\right|_{b_i \to b_i + 1} \;,
\nonumber \\
d_i & = & \Pi_{j=1}^{p+1} (1+a_j-b_i) \; ,
\nonumber \\
s_i (x)  & = & \frac{\Pi_{j=1}^{p+1} (x + a_j) - d_i }{x+b_i-1}
\sum_{j=0}^{p+1} P^{(p+1)}_{p+1-j}(\{a_r\}) \frac{\left[ x^{j} \!-\! (1\!-\!b_i)^{j} \right]}{x\!-\!(1\!-\!b_i)}
\nonumber \\
& = &
\sum_{j=0}^p P^{(p+1)}_{p-j}(\{a_r\}) \sum_{k=0}^j x^{j-k}(1\!-\!b_i)^k \;.
\label{diff:oper:2}
\end{eqnarray}
There
$\left. \right|_{a \to a+1} $ means substitution of $a$ by $a+1$,
and the polynomials  $P^{(p)}_j(r_1,\cdots,r_p)$  are defined as
\begin{equation}
\prod_{k=1}^{p}(z+r_k) =
\sum_{j=0}^{p} P^{(p)}_{p-j}(r_1,\cdots,r_p) z^j
\equiv \sum_{j=0}^p P^{(p)}_{p-j}(\vec{r}) z^j
\equiv \sum_{j=0}^p P^{(p)}_{j}(\vec{r}) z^{p-j} \;.
\label{P}
\end{equation}
$P^{(p)}_s(\vec{r})$   is a polynomial of order $s$ with respect to the variables $r$, and
$$
P^{(p)}_0(\vec{r}) = 1 \;,
\quad
P^{(p)}_j(\vec{r}) =
\sum_{i_1,\cdots,i_r=1}^p \prod_{i_1 < \cdots < i_j } r_{i_1} \cdots r_{i_j} \;, \quad j=1, \cdots, p \;.
$$
For example,
$P^{(p)}_1(\vec{r}) =  \sum_{j=1}^p r_j$ and
$P^{(p)}_p(\vec{r}) =  \prod_{j=1}^p r_j$.
Keeping in mind that
$$
\prod_{i=1}^{p}(z+r_i) \prod_{j=p+1}^{p+k}(z+r_j)
=
\prod_{l=1}^{p+k}(z+r_l) \;,
$$
we find that these polynomials satisfy the following relations:
\begin{equation}
P^{(p+k)}_{p+k-j}(r_1,\cdots,r_p,q_1,\cdots,q_k) =
\sum_{n=0}^k P^{(p)}_{p+1-j-n}(r_1,\cdots,r_p) P^{(k)}_{n}(q_1,\cdots,q_k)  \;,
\end{equation}
where  $j = 1, \cdots, p-k$
and 
$
P^{(p)}_{p+k}(\vec{r}) = 0
$
.
In particular,
\begin{eqnarray}
P^{(p+1)}_{p+1-j}(\vec{r},f) & = &  P^{(p)}_{p+1-j}(\vec{r}) \!+\! f  P^{(p)}_{p-j}(\vec{r})  \;, 
\quad j = 1, \cdots, p \;,
\nonumber \\
P^{(p+1)}_{p+1-j}(\vec{r}_{p-1},q_1,q_2) & = &  \sum_{k=0}^2 P^{(p-1)}_{p+1-j-k}(\vec{r}) P^{(2)}_{k}(\vec{q})  \;, 
\quad j = 1, \cdots, p-1 \;,
\nonumber \\
P^{(p+1)}_{p+1-j}(\vec{r}_{p-2},q_1,q_2,q_3) & = &  \sum_{k=0}^3 P^{(p-2)}_{p+1-j-k}(\vec{r}) P^{(3)}_{k}(\vec{q})  
\;, \quad j = 1, \cdots, p-2 \;.
\nonumber
\end{eqnarray}
The differential reduction  has the form of a product of several differential step-up/step-down operators, 
$H^{\pm}_{b_k}$ and $B^{\pm}_{a_k}$, respectively:
\begin{equation}
F(\vec{a}+\vec{m}; \vec{b}+\vec{n}; z)  =
\left( H_{\{ a\}}^{\pm} \right)^{\sum_i m_i} \left( B_{\{ b\}}^{\pm} \right)^{\sum_j n_j} F(\vec{a}; \vec{b}; z) \;,
\label{general}
\end{equation}
so that the maximal power of $\theta$ in this expression is equal to $r \equiv \sum_i m_i+\sum_j n_j$.
Since the hypergeometric function  ${}_{p+1}F_{p}(\vec{a};\vec{b}; z)$ satisfies the differential
equation of order $p\!+\!1$ (see Eq.~(\ref{diff})):
\begin{eqnarray}
&&
(1-z) \theta^{p+1}
{}_{p+1}F_{p}(\vec{a};\vec{b}; z)
\nonumber \\ &&
=
\Biggl\{
\sum_{r=1}^{p}
\left[
z P^{(p+1)}_{p+1-r}(\{a_j\})
-
P^{(p)}_{p+1-r}(\{b_j\!-\!1\})
\right] \theta^r
+
z \Pi_{k=1}^{p+1} a_k
\Biggr\}
{}_{p+1}F_{p}(\vec{a};\vec{b}; z) \;,
\end{eqnarray}
it is possible to express all terms containing higher powers of the operator $\theta^k$, where $k \geq p+1$,
in terms of product of $\theta^j$ with $j \leq p$ and 
rational functions of parameters and argument $z$.

In this way, any function  ${}_{p+1}F_{p}(\vec{a}+\vec{m};\vec{b}+\vec{k}; z)$ is expressible in terms
of the basic function and its first $p$-derivative:
\begin{eqnarray}
\label{decomposition}
&& \hspace{-5mm}
{}_{p+1}F_{p}(\vec{a}+\vec{m};\vec{b}+\vec{k}; z)
 =
\\ && \hspace{-5mm}
\frac{1}{S(a_i,b_j,z)}
\Biggl \{
R_1(a_i,b_j,z)
+ R_2(a_i,b_j,z) \theta
+
\cdots
+
  R_{p+1}(a_i,b_j,z) \theta^p
\Biggr\}
{}_{p+1}F_{p}(\vec{a};\vec{b}; z) \;,
\nonumber
\end{eqnarray}
where
$m,k$ is the set of integer numbers, and
$S$ and $R_i$ are polynomials in the parameters $\{a_i\},\{b_j\}$, and $z$.

From Eq.~(\ref{diff:oper:1}) it follows that if one of the upper parameters
$a_j$ is equal to unity, then the application of the step-down operator
$B^{-}_{a_j}$ to the hypergeometric function ${}_{p+1}F_{p}$
will produce unity,
$
B^{-}_{1} {}_{p+1}F_{p}(1, \vec{a};\vec{b}; z) \equiv 1 \;.
$
Taking into account the explicit form of the step-down operator $B^{-}_{1}$,
\begin{eqnarray}
B^{-}_{1} & = &
\frac{1}{\Pi_{k=1}^p(b_k-1)}
\left[ \Pi_{j=1}^p (b_j-1)
+ \sum_{j=1}^p P^{(p)}_{p-j}(\{b_k\!-\!1\}) \theta^j
- z \Pi_{j=1}^p \left(\theta+a_j\right)
\right] \;,
\end{eqnarray}
we get the differential identity
\begin{eqnarray}
&&
\Biggl\{
\Pi_{j=1}^p (b_j\!-\!1)
\!-\!
z \Pi_{j=1}^p a_j
\!+\!
(1\!-\!z) \theta^p
\Biggr\}
{}_{p+1}F_{p}(1, \vec{a};\vec{b}; z)
\nonumber \\ &&
+
\Biggl\{
\sum_{j=1}^{p-1} \left[ P^{(p)}_{p-j}(\{b_k-1\}) - z P^{(p)}_{p-j}(\{a_k\}) \right] \theta^j
\Biggr\}
{}_{p+1}F_{p}(1, \vec{a};\vec{b}; z)
=
\Pi_{k=1}^p(b_k\!-\!1) \;.
\label{relation}
\end{eqnarray}
The case when two or more upper parameters are equal to unity, $a_1=a_2=1$, does not generate
any new identities.

\section{Differential reduction of Appell hypergeometric functions}

\subsection{Appell hypergeometric functions: system of differential equations}
\label{Appell}
Let us consider the system of linear differential equations of  second order
for the functions $\omega(\vec{z})$:
\begin{eqnarray}
\theta_{11} \omega(\vec{z})
& = &
\Biggl\{
P_0(\vec{z}) \theta_{12}
+
P_1 (\vec{z}) \theta_1
+
P_2 (\vec{z}) \theta_2
+
P_3 (\vec{z})
\Biggr\}
\omega(\vec{z}) \;,
\label{eq1}
\\
\theta_{22} \omega(\vec{z})
& = &
\Biggl\{
R_0(\vec{z}) \theta_{12}
+
R_1 (\vec{z}) \theta_1
+
R_2 (\vec{z}) \theta_2
+
R_3 (\vec{z})
\Biggr\}
\omega(\vec{z}) \;,
\label{eq2}
\end{eqnarray}
where
$\vec{z} = (z_1,z_2)$
with
$z_1,z_2$ being variables,
$\{P_j,R_j\}$ are rational functions,
$\theta_j = z_j \partial_{z_j}$ for $j=1,2$,
and
$
\theta_{i_1\cdots i_k} = \theta_{i_i} \cdots \theta_{i_k}.
$
Using $\theta_j$ instead of the standard $\partial_j$
is explained by our applications.
Taking the derivative of Eq.(\ref{eq1}) with respect to $\theta_2$,  
using the well-known property
$
\partial_{2} \partial_{11} \omega(\vec{z})
=
\partial_{1} \partial_{12} \omega(\vec{z})
$
and applying Eq.~(\ref{eq2}), we rewrite Eq.~(\ref{eq1}) as follows:
\begin{eqnarray}
&&
\left[ \theta_1 \!-\! P_0 \theta_2 \right]
\theta_{12} \omega(\vec{z})
\nonumber \\ &&
 =
\Biggl\{
\left[
\theta_2 P_0
\!+\! P_1
\!+\! P_2 R_0
\right]
\theta_{12}
\!+\!
\left[
P_2 R_1
\!+\!
\theta_2 P_1
\right] \theta_1
\!+\!
\left[
P_2 R_2
\!+\!
\theta_2 P_2
\!+\! P_3
\right] \theta_2
\!+\!
P_2 R_3
\!+\!
\theta_2 P_3
\Biggr\}
\omega(\vec{z}) \;.
\nonumber \\ &&
\label{eq1a}
\end{eqnarray}
Applying a similar operation to  Eq.~(\ref{eq2}), we get
\begin{eqnarray}
&&
\left[
- R_0 \theta_1
\!+\! \theta_2
\right]
\theta_{12} \omega(\vec{z})
\nonumber \\ &&
 =
\Biggl\{
\left[
\theta_1 R_0
\!+\! R_2
\!+\! R_1 P_0
\right]
\theta_{12}
\!+\!
\left[
P_1 R_1
\!+\!
\theta_1 R_1
\!+\! R_3
\right] \theta_1
\!+\!
\left[
P_2 R_1
\!+\!
\theta_1 R_2
\right] \theta_2
\!+\!
R_1 P_3 \!+\! \theta_1 R_3
\Biggr\}
\omega(\vec{z}) \;.
\nonumber \\ &&
\label{eq2a}
\end{eqnarray}
It is well known \cite{appell} that under the condition
\begin{equation}
1 - P_0 R_0 \neq 0 \;,
\label{condition1}
\end{equation}
there are four independent solutions of the system of Eqs.~(\ref{eq1}) and (\ref{eq2}).
In this case, Eqs.~(\ref{eq1a}) and (\ref{eq2a}) can be solved, so that
\begin{eqnarray}
(1 \!-\! P_0 R_0 )
\theta_{112} \omega(\vec{z})
& = &
\Biggl\{
\left[
P_0
\left(
\theta_1 R_0
\!+\! R_2
\!+\! R_1 P_0
\right)
+
\theta_2 P_0
\!+\! P_1
\!+\! P_2 R_0
\right]
\theta_{12}
\nonumber \\ &&
\!+\!
\left[
P_2 R_1
\!+\!
\theta_2 P_1
\!+\!
P_0
\left(
P_1 R_1
\!+\!
\theta_1 R_1
\!+\! R_3
\right)
\right] \theta_1
\nonumber \\ &&
\!+\!
\left[
P_2 R_2
\!+\!
\theta_2 P_2
\!+\!
P_3
\!+\!
P_0
\left(
P_2 R_1
\!+\!
\theta_1 R_2
\right)
\right] \theta_2
\nonumber \\ &&
+
P_2 R_3
\!+\!
\theta_2 P_3
\!+\!
P_0
\left(
R_1 P_3 \!+\! \theta_1 R_3
\right)
\Biggr\}
\omega(\vec{z}) \;,
\label{eq1b}
\\
(1 \!-\! P_0 R_0 ) \theta_{122} \omega(\vec{z})
& = &
\Biggl\{
\left[
\theta_1 R_0
\!+\! R_2
\!+\!
R_1 P_0
\!+\!
R_0
\left(
\theta_2 P_0
\!+\! P_1
\!+\! P_2 R_0
\right)
\right]
\theta_{12}
\nonumber \\ &&
\!+\!
\left[
R_0
\left(
P_2 R_1
\!+\!
\theta_2 P_1
\right)
\!+\!
P_1 R_1
\!+\!
\theta_1 R_1
\!+\! R_3
\right] \theta_1
\nonumber \\ &&
\!+\!
\left[
P_2 R_1
\!+\!
\theta_1 R_2
\!+\!
R_0
\left(
P_2 R_2
\!+\!
\theta_2 P_2
\!+\!
P_3
\right)
\right] \theta_2
\nonumber \\ &&
+ R_0
\left(
P_2 R_3 \!+\! \theta_2 P_3
\right)
+
R_1 P_3 \!+\! \theta_1 R_3
\Biggr\}
\omega(\vec{z}) \;.
\label{eq2b}
\end{eqnarray}
The condition of complete integrability is defined via the relation
$
\partial_1 \left( \partial_{122} \omega(\vec{z}) \right)
=
\partial_2 \left( \partial_{112} \omega(\vec{z}) \right)
$.
In terms of the quantities in Eqs.~(\ref{eq1b}) and  (\ref{eq2b}),
it has the following form:
\begin{eqnarray}
&&
\theta_2 \left( \frac{1}{1-P_0 R_0} \left[\mbox{r.h.s.Eq.}~(\ref{eq1b}) \right]\right)
-
\theta_1 \left( \frac{1}{1-P_0 R_0} \left[\mbox{r.h.s.Eq.}~(\ref{eq2b}) \right]\right)
=
0 
\;,
\label{full}
\end{eqnarray}
where we have used the relation
$
\theta_{jj} = z_j^2 \partial_j^2 + z_j \partial_j
$.
If the condition of Eq.~(\ref{full}) is valid,  
then Eqs.~(\ref{eq1}) and (\ref{eq2})
can be reduced to the Pfaff system of  four differential equations
\begin{equation}
d \vec{f} = R \vec{f} \;,
\label{pfaff}
\end{equation}
where
$
\vec{f} = \left( \omega(\vec{z}), \theta_1 \omega(\vec{z}), \theta_2 \omega(\vec{z}), \theta_{12} \omega(\vec{z}) \right).
$

In the case
\begin{equation}
1 - P_0 R_0 = 0 \;,
\label{condition2}
\end{equation}
$\theta_{12} \omega(\vec{z})$ is expressible in terms of three other elements,
$
\omega(\vec{z}), \theta_1 \omega(\vec{z})
$ 
and  
$
\theta_2 \omega(\vec{z})
$.
In particular, using the notation of Eqs.~(\ref{eq1a}) and (\ref{eq2a}),
we have
\begin{eqnarray}
& &
\left[
\theta_2 P_0
+ P_1
+ P_2 R_0
+ P_0 \left(  \theta_1 R_0 + R_2 + R_1 P_0 \right)
\right]
\theta_{12}
\omega(\vec{z})
\nonumber \\
& = &
\Biggl\{
-
\left[
 P_0 \left(  \theta_1 R_1 + R_3 + R_1 P_1 \right)
+ P_2 R_1 + \theta_2 P_1
\right] \theta_1
\nonumber \\ &&
-
\left[
  P_0 \left(  \theta_1 R_2  + R_1 P_2 \right)
+ P_2 R_2 + \theta_2 P_2 + P_3
\right] \theta_2
\nonumber \\ &&
-
\left[
 P_0 \left(  \theta_1 R_3  + R_1 P_3 \right)
+ P_2 R_3 + \theta_2 P_3
\right]
\Biggr\} \omega(\vec{z}) \;.
\label{aux}
\end{eqnarray}
In this case, the integrability conditions are valid,
Eqs.~(\ref{eq1}) and (\ref{eq2})
can be reduced to the Pfaff system of Eq.~(\ref{pfaff}) of three differential equations
$
\vec{f} = \left( \omega(\vec{z}), \theta_1 \omega(\vec{z}), \theta_2 \omega(\vec{z}) \right),
$
and the system has three solutions.

For the Appell hypergeometric functions $F_1,F_2,F_3,$ and $F_4,$
the values of the coefficients in Eqs.~(\ref{eq1}) and (\ref{eq2})
are collected in Table~\ref{tab:1}.

\begin{table}
\caption{
Values of the coefficients in Eqs.~(\ref{eq1}) and (\ref{eq2}) for the Appell hypergeometric functions 
$F_1,F_2,F_3$ and $F_4$.}
\label{tab:1}
$$
\begin{tabular}[h]{|c|c|c|c|c|}
\hline
     &  $F_1$ & $F_2$ & $F_3$ & $F_4$\\
\hline
$P_0$  & - 1  & $\frac{z_1}{1-z_1}$ & $ - \frac{1}{(1-z_1)}$  &   $ \frac{2 z_1}{(1-z_1-z_2)}$  \\
\hline
$R_0$  & - 1  & $\frac{z_2}{1-z_2}$ & $ - \frac{1}{(1-z_2)}$  &   $ \frac{2 z_2}{(1-z_1-z_2)}$  \\
\hline
$P_1$  & $\frac{(a+b_1)z_1-(c-1)}{(1-z_1)}$  &  $\frac{(a+b_1)z_1 - (c_1-1)}{(1-z_1)}$
      &  $\frac{(a_1+b_1)z_1 - (c-1)}{(1-z_1)}$ & $\frac{(a+b)z_1-(c_1-1)(1-z_2)}{(1-z_1-z_2)}$ \\
\hline
$R_1$  & $\frac{ b_2 z_2}{(1-z_2)}$   & $\frac{b_2 z_2}{(1-z_2)}$ & 0 & $\frac{(a+b+1-c_1)z_2}{(1-z_1-z_2)}$\\
\hline
$P_2$  &  $\frac{ b_1 z_1}{(1-z_1)}$  & $\frac{b_1 z_1 }{(1-z_1)}$
       & 0  & $\frac{(a+b+1-c_2)z_1}{(1-z_1-z_2)}$ \\
\hline
$R_2$  &  $\frac{(a+b_2)z_2-(c-1)}{(1-z_2)}$  & $\frac{(a+b_2)z_2 - (c_2-1)}{(1-z_2)}$
       & $\frac{(a_2+b_2)z_2 - (c-1)}{(1-z_2)}$ & $\frac{(a+b)z_2-(c_2-1)(1-z_1)}{(1-z_1-z_2)}$ \\
\hline
$P_3$  &  $\frac{a b_1 z_1}{(1-z_1)}$   & $\frac{ab_1 z_1 }{(1-z_1)}$ & $\frac{a_1 b_1 z_1}{(1-z_1)}$ & $\frac{abz_1}{(1-z_1-z_2)}$ \\
\hline
$R_3$  &  $\frac{a b_2 z_2}{(1-z_2)}$   &  $\frac{ab_2 z_2}{(1-z_2)}$ & $\frac{a_2 b_2 z_2}{(1-z_2)}$ & $\frac{abz_2}{(1-z_1-z_2)}$ \\
\hline
\end{tabular}
$$
\end{table}

\subsection{Appell hypergeometric function $F_1$ }
\label{f1}
\subsubsection{General consideration}
Let us consider the Appell hypergeometric function $F_1$ defined around $x=y=0$ as
\begin{eqnarray}
\omega\equiv F_1(a, b_1, b_2, c; x,y)
=
\sum_{m=0}^\infty
\sum_{n=0}^\infty
\frac{(a)_{m+n} (b_1)_m (b_2)_n}{(c)_{m+n}}
\frac{x^m}{m!}
\frac{y^n}{n!} \;.
\label{definition:f1}
\end{eqnarray}
In this case,  Eqs.~(\ref{eq1}) and (\ref{eq2}) have the following form:
\begin{eqnarray}
\theta_{xx} \omega & = &
\left[
\frac{(a+b_1) x \!-\! (c-1)}{1-x} - b_2 \frac{y}{x-y}
\right] \theta_x  \omega
+ \frac{b_1 x (1-y)}{(1-x)(x-y)} \theta_y \omega
+ \frac{x}{1-x} a b_1 \omega \;,
\\
\theta_{yy} \omega & = &
\left[
\frac{(a+b_2) y \!-\! (c-1)}{1-y} + b_1 \frac{x}{x-y}
\right] \theta_y \omega
-  \frac{b_2 y (1-x)}{(1-y)(x-y)} \theta_x \omega
+ \frac{y}{1-y} a b_2 \omega \;.
\label{diff:F1}
\end{eqnarray}
Eq.~(\ref{condition2}) is fulfilled, and Eq.~(\ref{aux}) has the following form:
 \begin{equation}
(x-y) \frac{\partial^2 \omega}{\partial x \partial y}
- b_2 \frac{\partial \omega}{\partial x}
+ b_1 \frac{\partial \omega}{\partial y}
= 0 \;,
\end{equation}
or in terms of the operators $\theta_x, \theta_y$:
\begin{eqnarray}
\theta_{xy} \omega  =
\frac{b_2 y}{x-y} \theta_x \omega
-
\frac{b_1 x}{x-y} \theta_y \omega
\;.
\label{f1:aux}
\end{eqnarray}
\subsubsection{Differential reduction of $F_1$}

The direct differential expressions follow from Eqs.~(\ref{do1}) and (\ref{do2}), 
\begin{eqnarray}
a F_1(a + {\bf 1},b_1,b_2,c;x,y) & = &
  (\theta_x \!+\! \theta_y \!+\! a ) F_1(a,b_1,b_2,c;x,y) \;,
\label{directF1:a}
\\ \vspace{10mm}
b_1 F_1(a,b_1+ {\bf 1},b_2,c;x,y) & = &
  (\theta_x \!+\! b_1 ) F_1(a,b_1,b_2,c;x,y)  \;,
\label{directF1:b1}
\\ \hspace{15mm}
(c-1) F_1(a,b_1,b_2,c - {\bf 1};x,y) & = &
  (\theta_x \!+\! \theta_y \!+\! c \!-\! 1 ) F_1(a,b_1,b_2,c;x,y) \;.
\label{directF1:c}
\end{eqnarray}
The inverse differential relations were considered in Refs.~\cite{appell,mullen}:
\begin{eqnarray}
&&
(c \!-\! a) F_1(a - {\bf 1},b_1,b_2,c;x,y)  =
\nonumber \\ &&
\left[
c \!-\! a \!-\! b_1 x \!-\! b_2 y
\!+\! (1\!-\!x) \theta_x
\!+\! (1\!-\!y) \theta_y
\right]
F_1(a,b_1,b_2,c;x,y) \;,
\label{inverseF1:a}
\\ &&
(c\!-\! b_1 \!-\! b_2)
F_1(a,b_1 - {\bf 1},b_2,c;x,y)  =
\nonumber \\ &&
\left[
c \!-\! b_1 \!-\! b_2 \!-\! ax
\!+\! (1\!-\!x) \theta_x
\!-\! x \left(1\!-\!\frac{1}{y} \right) \theta_y
\right]
F_1(a,b_1,b_2,c;x,y)  \;,
\label{inverseF1:b1}
\\ &&
(c\!-\!a) (c \!-\! b_1 \!-\! b_2)F_1(a,b_1,b_2,c + {\bf 1};x,y)
=
\nonumber \\ &&
c
\left[
  (c\!-\! a \!-\! b_1 \!-\! b_2)
\!-\! \left( 1 \!-\! \frac{1}{x} \right) \theta_x
\!-\! \left( 1 \!-\! \frac{1}{y} \right) \theta_y
\right]
F_1(a,b_1,b_2,c;x,y) \;.
\label{inverseF1:c}
\end{eqnarray}
The differential reduction for the parameter $b_2$ follows from Eqs.~(\ref{directF1:b1}) and (\ref{inverseF1:b1}),
and the symmetry property of the function $F_1$,
$
F_1(a,b_1,b_2,c;x,y) =  F_1(a,b_2,b_1,c;y,x) \;,
$
i.e.
$$
b_1 \Leftrightarrow b_2 \;, \quad
x  \Leftrightarrow y  \;.
$$
\subsection{Appell hypergeometric function $F_2$}
\label{f2}
\subsubsection{General consideration}
Let us consider the Appell hypergeometric function $F_2$ defined around $x=y=0$ as
\begin{eqnarray}
\omega \equiv F_2(a, b_1, b_2, c_1, c_2 ; x,y)
=
\sum_{m=0}^\infty
\sum_{n=0}^\infty
\frac{(a)_{m+n} (b_1)_m (b_2)_n}{(c_1)_m (c_2)_n}
\frac{x^m}{m!}
\frac{y^n}{n!} \;,
\label{definition:f2}
\end{eqnarray}
In this case  Eqs.~(\ref{eq1}) and  (\ref{eq2}) have the following form:
\begin{eqnarray}
(1 \!-\! x) \theta_{xx} \omega
& = & x \theta_{xy} \omega
+
\left[
(a \!+\! b_1)x \!-\! (c_1\!-\!1)
\right] \theta_x \omega
\!+\! b_1 x \theta_y \omega
\!+\! a b_1 x \omega  \;,
\\
(1\!-\!y) \theta_{yy} \omega
& = & y \theta_{xy} \omega
+
\left[
(a \!+\! b_2)y \!-\! (c_2 \!-\! 1)
\right] \theta_y \omega
\!+\! b_2 y  \theta_x \omega
\!+\! a b_2 y \omega \;.
\label{f2:diff}
\end{eqnarray}
The condition of Eq.~(\ref{condition1}) is fulfilled, and Eqs.~(\ref{eq1b}) and (\ref{eq2b}) have the following form:
\begin{eqnarray}
(1\!-\!x\!-\!y) \theta_{xxy} \omega  & = &
\left[ \left( a \!+\! b_1 \!+\! 1 \!-\! c_2 \right)x \!-\! (c_1 \!-\! 1) (1 \!-\! y) \!+\! \frac{b_2 xy }{1\!-\!x} \right]
\theta_{xy} \omega
\nonumber \\ &&
\!+\! \frac{b_2 xy}{1\!-\!x} \left[ a \!+\! b_1 \!+\! 1 \!-\! c_1 \right] \theta_x \omega
\nonumber \\ &&
+ \left[ (a \!+\! 1 \!-\! c_2 ) b_1 x \!+\! \frac{b_1 b_2 xy}{1\!-\!x} \right] \theta_y \omega
\nonumber \\ &&
+
\frac{a b_1 b_2 x y}{1\!-\!x} \omega  \;,
\\
(1\!-\!x\!-\!y) \theta_{yyx} \omega  & = &
\left[ \left( a \!+\! b_2 \!+\! 1 \!-\! c_1 \right) y \!-\! (c_2 \!-\! 1) (1\!-\!x) \!+\! \frac{b_1 xy }{1\!-\!y} \right]
\theta_{xy} \omega
\nonumber \\ &&
+ \frac{b_1 xy}{1\!-\!y} \left[ a \!+\! b_2 \!+\! 1 \!-\! c_2 \right] \theta_y \omega
\nonumber \\ &&
+ \left[ (a \!+\! 1 \!-\! c_1 ) b_2 y \!+\! \frac{b_1 b_2 xy}{1\!-\!y} \right] \theta_x \omega
\nonumber \\ &&
+
\frac{a b_1 b_2 x y}{1\!-\!y} \omega  \;.
\end{eqnarray}
\subsubsection{Differential reduction of $F_2$}
The direct differential expressions  follow from Eqs.~(\ref{do1}) and (\ref{do2}),
\begin{eqnarray}
a F_2(a+{\bf 1},b_1,b_2,c_1,c_2;x,y)  & =  &
(a \!+\! \theta_x \!+\! \theta_y) F_2(a,b_1,b_2,c_1,c_2;x,y) \;,
\label{directF2:a}
\\
b_1 F_2(a,b_1+{\bf 1},b_2,c_1,c_2;x,y) & = &
(b_1 \!+\! \theta_x) F_2(a,b_1,b_2,c_1,c_2;x,y) \;,
\label{directF2:b1}
\\
(c_1 \!-\! 1) F_2(a,b_1,b_2,c_1-{\bf 1},c_2;x,y) & = &
(c_1 \!-\! 1 \!+\! \theta_x) F_2(a,b_1,b_2,c_1,c_2;x,y) \;,
\label{directF2:c1}
\end{eqnarray}
The inverse differential relations were considered in Ref.~\cite{mullen}:
\begin{eqnarray}
&&
F_2(a-{\bf 1},b_1,b_2,c_1,c_2;x,y)
=
\Biggl\{
1
\!-\! \frac{xb_1 \!-\! (1\!-\!x) \theta_x}{c_1\!-\!a}
\!-\! \frac{yb_2 \!-\! (1\!-\!y) \theta_y}{c_2\!-\!a}
\nonumber \\ &&
+ \frac{1}{c_1 \!+\! c_2 \!-\! a \!-\! 1} \left[ \frac{1}{c_1 \!-\! a} \!+\! \frac{1}{c_2 \!-\! a} \right]
\left[
(1\!-\!x\!-\!y) \theta_{xy} \!-\! b_1 x \theta_y \!-\! b_2 y \theta_x
\right]
\Biggr\}
\nonumber \\ &&
\vspace{5mm}
\times
F_2(a,b_1,b_2,c_1,c_2;x,y)
\;,
\label{inverseF2:a}
\\ &&
F_2(a,b_1-{\bf 1},b_2,c_1,c_2;x,y)
=
\Biggl\{
1 + \frac{ (1\!-\!x) \theta_x \!-\! x (a \!+\! \theta_y) }{c_1\!-\!b_1}
\Biggr\}
F_2(a,b_1,b_2,c_1,c_2;x,y)
\;,
\label{inverseF2:b1}
\\  &&
F_2(a,b_1,b_2,c_1+{\bf 1},c_2;x,y)
=
\frac{c_1}{(c_1\!-\!a)(c_1\!-\!b_1)}
\Biggl\{
c_1 \!-\! a \!-\! b_1 \!-\! \left( 1 \!-\! \frac{1}{x} \right) \theta_x
\nonumber \\ &&
\!-\! \frac{1}{x(c_1\!+\!c_2\!-\!a\!-\!1)}
\left[
x b_1 \theta_y \!+\! y b_2 \theta_x \!-\! (1\!-\!x\!-\!y) \theta_{xy}
\right]
\Biggr\}
F_2(a,b_1,b_2,c_1,c_2;x,y)
\;.
\label{inverseF2:c1}
\end{eqnarray}
The differential reductions for the parameters $b_2$ and $c_2$ follow from
Eqs.~(\ref{directF2:b1}),(\ref{inverseF2:b1}), 
and Eqs.~(\ref{directF2:c1}),(\ref{inverseF2:c1}), respectively, and
the symmetry property of the function $F_2$,
$
F_2(a,b_1,b_2,c;x,y) = F_2(a,b_2,b_1,c;y,x),
$
i.e.
$$
b_1 \Leftrightarrow b_2 \;, \quad
c_1 \Leftrightarrow c_2 \;, \quad
x  \Leftrightarrow y  \;.
$$
\subsection{Appell hypergeometric function $F_3$}
\label{f3}
\subsubsection{General consideration}
Let us consider the Appell hypergeometric function $F_3$ defined around $x=y=0$ as
\begin{eqnarray}
\omega \equiv F_3(a_1, a_2, b_1, b_2, c; x,y)
=
\sum_{m=0}^\infty
\sum_{n=0}^\infty
\frac{(a_1)_{m} (a_2)_{n} (b_1)_m (b_2)_n}{(c)_{m+n} }
\frac{x^m}{m!}
\frac{y^n}{n!} \;,
\label{definition:f3}
\end{eqnarray}
In this case,  Eqs.~(\ref{eq1}) and  (\ref{eq2}) have the following form:
\begin{eqnarray}
(1\!-\!x) \theta_{xx}  \omega
& = &  - \theta_{xy} \omega
+
\left[
(a_1 \!+\! b_1) x \!-\! (c \!-\! 1)
\right] \theta_x \omega
\!+\! x a_1 b_1 \omega \;,
\\
(1\!-\!y) \theta_{yy} \omega
& = &  - \theta_{xy} \omega
+
\left[
(a_2 \!+\! b_2) y \!-\! (c \!-\!1 )
\right] \theta_y \omega
\!+\! y a_2 b_2 \omega \;.
\end{eqnarray}
The condition of Eq.~(\ref{condition1}) is fulfilled, and Eq.~(\ref{eq1b}) and (\ref{eq2b}) have the following form:
\begin{eqnarray}
(xy\!-\!x\!-\!y) \theta_{xxy} \omega & = &
\left[
(1\!-\!y)(a_1 \!+\! b_1)x \!-\! y (a_2 \!+\! b_2 \!+\! 1 \!-\! c)
\right] \theta_{xy} \omega
\nonumber \\ &&
+ (1\!-\!y) x a_1 b_1 \theta_y \omega
\!-\! y a_2 b_2 \theta_x \omega \;,
\\
(xy\!-\!x\!-\!y) \theta_{xyy} \omega & = &
\left[
(1\!-\!x)(a_2 \!+\! b_2)y \!-\! x (a_1 \!+\! b_1 \!+\! 1 \!-\! c)
\right] \theta_{xy} \omega
\nonumber \\ &&
+ (1\!-\!x) y a_2 b_2 \theta_x \omega
\!-\! x a_1 b_1 \theta_y \omega\;.
\end{eqnarray}
\subsubsection{Differential reduction of $F_3$}
The direct differential expressions follow from Eqs.~(\ref{do1}) and (\ref{do2}),
\begin{eqnarray}
a_1 F_3(a_1+{\bf 1},a_2,b_1,b_2,c;x,y)  & =  &
(a_1 \!+\! \theta_x ) F_3(a_1,a_2,b_1,b_2,c;x,y) \;,
\label{directF3:a1}
\\
b_1 F_3(a_1,a_2,b_1+{\bf 1},b_2,c;x,y) & = &
(b_1 \!+\! \theta_x) F_3(a_1,a_2,b_1,b_2,c;x,y) \;,
\label{directF3:b1}
\\
(c \!-\! 1) F_3(a_1,a_2,,b_1,b_2,c-{\bf 1};x,y) & = &
(c \!-\! 1 \!+\! \theta_x \!+\! \theta_y) F_3(a,b_1,b_2,c_1,c_2;x,y) \;.
\label{directF3:c}
\end{eqnarray}
The inverse differential relations were considered in Ref.~\cite{singal}:
\begin{eqnarray}
&&
F_3(a_1-{\bf 1},a_2,b_1,b_2,c;x,y)
= 1
+
\frac{1}{(c\!-\!a_1\!-\!a_2)(c\!-\!b_2\!-\!a_1)}
\nonumber \\ &&
\Biggl\{
(c \!-\! b_2 \!-\! a_1 \!-\! a_2) \left[
(1 \!-\! x) \theta_x \!-\! x b_1
\right]
\!+\!
b_1 x \left( 1 \!-\! \frac{1}{y} \right) \theta_y
\!-\!
\left( 1 \!-\! x \!+\! \frac{x}{y} \right) \theta_{xy}
\Biggr\}
\nonumber \\ &&
\hspace{15mm}
\times
F_3(a,b_1,b_2,c_1,c_2;x,y) \;,
\label{inverseF3:a1}
\\ &&
F_3(a_1,a_2,b_1-{\bf 1},b_2,c;x,y)
= 1
+
\frac{1}{(c \!-\! b_1 \!-\! b_2)(c\!-\! a_2 \!-\! b_1)}
\nonumber \\ &&
\Biggl\{
(c \!-\! a_2 \!-\! b_1 \!-\! b_2) \left[
(1 \!-\! x) \theta_x \!-\! x a_1
\right]
+
a_1 x \left( 1 \!-\! \frac{1}{y} \right) \theta_y
-
\left( 1 \!-\! x \!+\! \frac{x}{y} \right) \theta_{xy}
\Biggr\}
\nonumber \\ &&
\hspace{15mm}
\times
F_3(a,b_1,b_2,c_1,c_2;x,y) \;,
\label{inverseF3:b1}
\\ &&
\Delta
F_3(a_1,a_2,b_1,b_2,c+{\bf 1};x,y)
=
\nonumber \\ &&
\hspace{5mm}
c
\Biggl\{
A
-
D_1
\left( 1 \!-\! \frac{1}{x} \right) \theta_x
-
D_2
\left( 1 \!-\! \frac{1}{y} \right) \theta_y
+
B
\left( 1 \!-\! \frac{1}{x} \!-\! \frac{1}{y} \right) \theta_{xy}
\Biggr\}
\nonumber \\ &&
\hspace{15mm}
\times
F_3(a,b_1,b_2,c_1,c_2;x,y)
\;,
\label{inverseF3:c}
\end{eqnarray}
where in the last expression,
\begin{eqnarray}
&&
\delta_1 = c \!-\! a_1 \!-\! b_1 \;, \quad
\delta_2 = c \!-\! a_2 \!-\! b_2 \;,  \quad
F =  c \!-\! a_1 \!-\! a_2 \!-\! b_1 \!-\! b_2 \;,
\nonumber \\ &&
A = \delta_1 \delta_2 F \!+\! a_1 b_1 \delta_1 \!+\! a_2 b_2 \delta_2 \;,
\nonumber \\ &&
D_1 =  \delta_2 F \!+\! a_1 b_1 \!-\! a_2 b_2 \;, \quad
D_2 =  \delta_1 F \!+\! a_2 b_2 \!-\! a_1 b_1 \;,  \quad
B = \delta_1 + \delta_2 \;,
\nonumber \\ &&
\Delta =
(c \!-\! b_1  \!-\! b_2) (c \!-\! a_1 \!-\! a_2) (c \!-\! a_2 \!-\! b_1) (c \!-\! a_1 \!-\! b_2)
\;.
\end{eqnarray}
The differential reductions for the parameters $a_2$ and $b_2$ follow from
Eqs.~(\ref{directF3:a1}), (\ref{inverseF3:a1})
and Eqs.~(\ref{directF3:b1}), (\ref{inverseF3:b1}), respectively,
and the symmetry property of the function $F_3,$

$
F_3(a_1,a_2,b_1,b_2,c;x,y)
=
F_3(a_2,a_1,b_2,b_1,c;y,x) \;,
$
i.e.
$$
a_1 \Leftrightarrow a_2 \;, \quad
b_1 \Leftrightarrow b_2 \;, \quad
x  \Leftrightarrow y  \;.
$$
\subsection{Appell hypergeometric function $F_4$}
\label{f4}
\subsubsection{General consideration}
Let us consider the Appell hypergeometric function $F_4$ defined around $x=y=0$ as
\begin{eqnarray}
\omega \equiv F_4(a, b, c_1, c_2; x,y)
=
\sum_{m=0}^\infty
\sum_{n=0}^\infty
\frac{(a)_{m+n} (b)_{m+n} }{(c_1)_m (c_2)_n}
\frac{x^m}{m!}
\frac{y^n}{n!} \;.
\label{definition:f4}
\end{eqnarray}
The condition of Eq.~(\ref{condition1}) is fulfilled, and Eqs.~(\ref{eq1b}),(\ref{eq2b}) have the following form:
\begin{eqnarray}
(1\!-\!x\!-\!y) \theta_{xx} \omega
& = &
2 x \theta_{xy} \omega
+
\left[
(a\!+\!b)x \!-\! (c_1\!-\!1)(1\!-\!y)
\right] \theta_x \omega
+
\left[
a\!+\!b \!+\! 1 \!-\! c_2
\right] x \theta_y \omega
\!+\! 
a b x  \omega \;,
\label{F4:xx}
\\
(1\!-\!x\!-\!y) \theta_{yy} \omega
& = &
2 y \theta_{xy} \omega
+
\left[
(a\!+\!b)y \!-\! (c_2\!-\!1)(1\!-\!x)
\right] \theta_y \omega
+
\left[
a\!+\!b \!+\! 1 \!-\! c_1
\right] y \theta_x \omega
\!+\! 
a b y  \omega \;.
\label{F4:yy}
\end{eqnarray}
The condition of Eq.~(\ref{condition1}) is fulfilled.
However, instead of  Eqs.~(\ref{eq1b}) and (\ref{eq2b}), which are very lengthy in this case, we present the results
in the following form:
%
%
%

\begin{eqnarray}
&&
\left[(1 \!-\! x \!-\! y )^2 \!-\! 4xy \right] \theta_{xxy} \omega =
\nonumber \\ &&
y \left[
1 \!-\! x \!-\! y \!+\! 2x( a \!+\! b \!+\! 1 \!-\! c_1)
\right]
\theta_{xx} \omega
+
x \left[
2 x \!+\! (1 \!-\! x \!-\! y) ( a \!+\! b \!+\! 1 \!-\! c_2)
\right]\theta_{yy} \omega
+
\nonumber \\ &&
\Biggl[
2 (a \!+\! b \!+\! 1 \!-\! c_2) (1\!-\!x) x
- (c_1 \!-\! 1) (1\!-\!x\!-\!y)^2
- x (1 \!-\! x \!-\! y) ( a \!+\! b \!+\! c_1 \!-\! 1)
\Biggr] \theta_{xy} \omega
\nonumber \\ &&
+
y \left[
(c_1 \!-\! 1)( 1 \!-\! x \!-\! y)
+ 2 a b x
\right]
\theta_{x} \omega
+
x \left[
2 x (c_2 \!-\! 1)
+
( 1 \!-\! x \!-\! y) a b
\right]
\theta_{y} \omega
\;,
\\[10mm] &&
\left[(1 \!-\! x \!-\! y )^2 \!-\! 4xy \right] \theta_{yyx} \omega =
\nonumber \\ &&
x \left[
1 \!-\! x \!-\! y \!+\! 2y( a \!+\! b \!+\! 1 \!-\! c_2)
\right]
\theta_{yy} \omega
+
y \left[
2 y \!+\! (1 \!-\! x \!-\! y) ( a \!+\! b \!+\! 1 \!-\! c_1)
\right]\theta_{xx} \omega
+
\nonumber \\ &&
\Biggl[
2 (a \!+\! b \!+\! 1 \!-\! c_1) (1\!-\!y) y
- (c_2 \!-\! 1) (1\!-\!x\!-\!y)^2
- y (1 \!-\! x \!-\! y) ( a \!+\! b \!+\! c_2 \!-\! 1)
\Biggr] \theta_{xy} \omega
\nonumber \\ &&
+
x \left[
(c_2 \!-\! 1)( 1 \!-\! x \!-\! y)
+ 2 a b y
\right]
\theta_{y} \omega
+
y
\left[
2 y (c_1 \!-\! 1)
+
( 1 \!-\! x \!-\! y) a b
\right]
\theta_{x} \omega
\;,
\end{eqnarray}
where the values of
$\theta_{xx} \omega$
and
$\theta_{yy} \omega$
are taken from Eqs.~(\ref{F4:xx}) and (\ref{F4:yy}).
\subsubsection{Differential reduction of $F_4$}
The direct differential expressions follow from Eqs.~(\ref{do1}) and (\ref{do2}),
\begin{eqnarray}
a F_4(a+{\bf 1},b,c_1,c_2;x,y)  & =  &
(a \!+\! \theta_x \!+\! \theta_y) F_4 (a,b,c_1,c_2;x,y)  \;,
\label{directF4:a}
\\
(c_1 \!-\! 1) F_4(a,b,c_1-{\bf 1},c_2;x,y) & = &
(c_1 \!-\! 1 \!+\! \theta_x ) F_4(a,b,c_1,c_2;x,y) \;.
\label{directF4:c1}
\end{eqnarray}
The inverse differential relations were considered in Ref.~\cite{mullen}:
\begin{eqnarray}
&&
F_4(a\!-\!{\bf 1},b,c_1,c_2;x,y)
=
\Biggl\{
1
\nonumber \\ &&
-
\frac{x}{(c_1\!-\!a)}
\left[
\left( 1\!-\!\frac{1}{x} \right) \theta_x \!+\! \theta_y \!+\! b
\right]
\!-\!
\frac{y}{(c_2\!-\!a)}
\left[
\left( 1\!-\!\frac{1}{y} \right)\theta_y  \!+\! \theta_x \!+\! b
\right]
\nonumber \\ &&
+
\frac{1}{(1-x-y)(c_1+c_2-a-1)}
\left[
\frac{1}{c_1\!-\!a}
\!+\!
\frac{1}{c_2\!-\!a}
\right]
\Biggl(
\nonumber \\ && \hspace{15mm}
[(1-x-y)^2-4 xy] \theta_{xy}
\nonumber \\ && \hspace{15mm}
- y [2 x (a \!+\! b \!+\! 1 \!-\! c_1 ) \!+\! (1\!-\!x\!-\!y)(b\!+\!1\!-\!c_1)] \theta_x
\nonumber \\ && \hspace{15mm}
- x [2 y (a \!+\! b \!+\! 1 \!-\! c_2 ) \!+\! (1\!-\!x\!-\!y)(b\!+\!1\!-\!c_2)] \theta_y
\nonumber \\ && \hspace{15mm}
- 2 a b xy
\hspace{25mm}
\Biggl)
\Biggr\}
F_4(a,b,c_1,c_2;x,y) \;,
\label{inverseF4:a}
\\ [15mm] &&
x (c_1\!-\!a)(c_1\!-\!b)
F_4(a,b,c_1+{\bf 1},c_2;x,y) =
\nonumber \\ &&
c_1
\Biggl\{
x (c_1-a-b)
+
(1\!-\!x\!-\! y) \theta_x
-
\frac{x}{c_1}
(a \!+\! b \!+\! 1 \!-\! c_2) \theta_y
\nonumber \\ &&
+
\frac{(2c_1\!+\!c_2\!-\!a\!-\!b\!-\!1)}{(1\!-\!x\!-\!y)(c_1\!+\!c_2\!-\!a\!-\!1) (c_1\!+\!c_2\!-\!b\!-\!1)}
\Biggl(
[(1-x-y)^2 - 4 xy ] \theta_{xy}
\nonumber \\ && \hspace{20mm}
-
y \left[
(1\!-\!x\!-\!y)(a \!+\! b \!-\! c_2 \!-\! 2 c_1 \!+\! 2)
+ 2 x (a \!+\! b \!-\! c_1 \!+\! 1)
\right]
\theta_x
\nonumber \\ && \hspace{20mm}
-
\frac{x}{c_1}
\left[
2 c_1 y (a\!+\!b\!+\!1\!-\!c_2)
+
(1\!-\!x\!-\!y) (c_2\!-\!a\!-\!1)(c_2\!-\!b\!-\!1)
\right]
\theta_y
\nonumber \\ && \hspace{20mm}
-
2 ab xy
\hspace{45mm}
\Biggr)
\Biggr\}
F_4(a,b,c_1,c_2;x,y) \;.
\label{inverseF4:c1}
\end{eqnarray}
The differential reduction for the parameters $b$ and $c_2$ follows from
Eqs.~(\ref{directF4:a}), (\ref{inverseF4:a})
and Eqs.~(\ref{directF4:c1}), (\ref{inverseF4:c1}), respectively,
and the symmetry property of the function $F_4,$

(i)
$
F_4(a,b,c_1,c_2,c;x,y)
=
F_4(b,a,c_1,c_2,c;x,y) \;:
\quad
a   \Leftrightarrow b \;,
$

(ii)
$
F_4(a,b,c_1,c_2,c;x,y)
=
F_4(a,b,c_2,c_1,c;y,x) \;:
\quad
c_1 \Leftrightarrow c_2 \;, \quad
x   \Leftrightarrow y  \;.
$

\subsection{Appell hypergeometric functions: exceptional values of parameters}
\label{exception}

As was explained in Section~\ref{Appell}, the differential-reduction algorithm
as applied  to Appell functions may be written symbolically as
\begin{eqnarray}
&&
R(x,y) F_1(\vec{A}+\vec{m}; x,y)
=
\left[
P_0(x,y)
+ P_1(x,y) \theta_x
+ P_2(x,y) \theta_y
\right] F_1(\vec{A}; x,y) \;,
\label{reductionf1}
\\ &&
S(x,y) F_j(\vec{A}+\vec{m}; x,y)
=
\left[
Q_0(x,y)
\!+\! Q_1(x,y) \theta_x
\!+\! Q_2(x,y) \theta_y
\!+\! Q_3(x,y) \theta_{xy}
\right] F_j(\vec{A}; x,y) \;,
\nonumber \\ &
\label{reductionf234}
\end{eqnarray}
where $j=2,3,4$,
$\vec{m}$ is a set of integers, $\vec{A}$ is a set of parameters,
$R,S,P_i,Q_i$ are some polynomials,
and
$\theta_x = x \partial_x (\theta_y = y \partial_y)$.

However, there is a special subset of values of parameters for which 
the results of the differential reduction,
Eqs.~(\ref{reductionf1}) and (\ref{reductionf234}), have simpler forms.
This set of exceptional values of parameters can be defined from  
(i) the condition that the hypergeometric function entering the l.h.s. of
Eqs.~(\ref{inverseF1:a})--(\ref{inverseF1:c}),
     (\ref{inverseF2:a})--(\ref{inverseF2:c1}),
     (\ref{inverseF3:a1})--(\ref{inverseF3:c}),
     (\ref{inverseF4:a}), and (\ref{inverseF4:c1}),
is expressible in terms
of  simpler hypergeometric functions (e.g. Gauss hypergeometric functions);
(ii) the condition that some of the coefficients entering 
the inverse differential relations  are equal to zero(infinity).

For the Appell hypergeometric functions $F_1,F_2,F_3,$ and $F_4$,
the exceptional sets of parameters  are listed in Table~\ref{tab:2}.

\begin{table}
\caption{
Exceptional set of parameters for the Appell hypergeometric functions 
$F_1,F_2,F_3$ and $F_4$.}
\label{tab:2}
$$
\begin{tabular}[h]{|c|c|}
\hline
$F_1$ & $\{a, b_1, b_2, c\!-\!a, c\!-\!b_1\!-\!b_2\} \in \mathbb{Z}$\\
\hline
$F_2$ & $\{ a, b_1, b_2, c_1\!-\!a, c_2\!-\!a, c_1\!+\!c_2\!-\!a, c_1\!-\!b_1, c_2\!-\!b_2 \}\in \mathbb{Z}$ \\
\hline
$F_3$ & $\{a_1, a_2, b_1 ,b_2, c\!-\!a_1\!-\!a_2, c\!-\!b_1\!-\!b_2, c\!-\!a_2\!-\!b_1, c\!-\!a_1\!-\!b_2\}$  $\in \mathbb{Z}$\\
\hline
$F_4$ & $\{a, b, c_1\!-\!a, c_1\!-\!b, c_2\!-\!a,  c_2 \!-\! b, c_1\!+\!c_2\!-\!a, c_1\!+\!c_2\!-\!b \} \in \mathbb{Z}$ \\
\hline
\end{tabular}
$$
\end{table}
It is not surprising that the set of exceptional values of parameters
coincides with the set of parameters defining the condition of
irreducibility of the monodromy group of the corresponding hypergeometric functions
(see Ref.~\cite{monodromy} and references therein).
The condition of irreducibility of Mellin-Barnes integrals \cite{beukers}
is related to the criterion of irreducibility of  Feynman diagrams \cite{KK2012}. 

\section{pfq - differential reduction of hypergeometric function ${}_pF_{p-1}$.}
\label{HYPERDIRE}

\subsection{Non-exceptional values of parameters}
In this section, we will present the Mathematica\footnote{It was tested for Mathematica $7.0$.} based
package {\bf pfq} for the differential reduction of the hypergeometric function ${}_pF_{p-1}$.
In contrast to the version presented in Ref.~\cite{our:hep},
the current version  deals  with non-exceptional and exceptional values of parameters.
The Takayama  algorithm is implemented in the file {\bf pfq.m}, and an example of its application
is given in the file {\bf example-pfq.m}.

The program may be loaded in the standard way:
$$
 << \mathrm{"pfq.m"}
$$
and includes two routines:
{\bf ToGroebnerBasis}[$\dots$]
and
{\bf explicitForm}[$\dots$].

\noindent
The main routine,
\begin{equation}
{\bf ToGroebnerBasis}[\mbox {Argumentsvector}],
\label{ToGroebnerBasis}
\end{equation}
calculates explicitly the ratio of the functions $\{R_k \}$ and $S$ of Eq.~(\ref{decomposition}).
The
``{\rm Argumentsvector}''
in Eq.~(\ref{ToGroebnerBasis}) is the set of parameters of the hypergeometric function
on the l.h.s of Eq.~(\ref{decomposition}) with an explicit  set of integer numbers,
by which each parameter should be shifted:
\begin{equation}
\mbox{Argumentsvector} = \{\{\vec{a} + \vec{M} \}, \{\vec{b}+\vec{K} \}, \{x \}\} \;,
\end{equation}
where $\vec{M}, \vec{K}$ are integers, $\vec{a},\vec{b}$ are any symbols,
and $x$ denotes the argument of the hypergeometric function.
The output of {\bf ToGroebnerBasis}[\dots] has the following structure:
\begin{gather}
\{\{Q_1,\cdots,Q_{p+1}\},\{\{a_1,\cdots,a_{p+1}\},\{1+b_1,\cdots,1+b_p\},x\},\mbox {factor} \},
\label{output:1}
\end{gather}
where
\begin{itemize}
\item
$\{\{a_1,\cdots,a_{p+1}\},\{1+b_1,\ldots,1+b_p\},x\}$ are the set of parameters and the argument of the resulting 
hypergeometric function 
entering in the r.h.s. of Eq.~(\ref{decomposition});
\item
$\{Q_1,\cdots,Q_{p+1}\}$ are the rational functions whose numbering corresponds to the 
power of $\theta^{i}$, $i=0,\ldots,p$;
\item
$\mbox{factor}$ is an overall factor.
\end{itemize}
The explicit form of Eqs.~(\ref{ToGroebnerBasis})-(\ref{output:1}) is the following:
\begin{eqnarray}
&&
{}_{p+1}F_{p}\left(\begin{array}{c|}
 a_{1} \!+\! M_1, \cdots,  a_{p+1} \!+\! M_{p+1} \\
 b_{1} \!+\! K_1, \cdots,  b_{p} \!+\! K_p   \end{array} ~x \right)
\nonumber \\ &&
=
\mbox{factor}
\times
\biggl( Q_1 \!+\! Q_2\theta \!+\! \cdots \!+\! Q_{p+1} \theta^{p} \biggr)
{}_{p+1}F_{p}\left(\begin{array}{c|}
  a_{1}, \cdots,  a_{p+1} \\
 1\!+\!b_{1} , \cdots, 1\!+\!b_{p}   \end{array} ~x \right) \;.
\label{basis}
\end{eqnarray}

\begin{enumerate}
\item
{\bf Example 1}
\footnote{All functions in the package HYPERDIRE generate output without additional simplification.
This is done for the maximum efficiency of the algorithm.
To bring the output into  a simpler form, we recommend to use in addition the command 
{\bf Simplify}. In particular, all considered examples
are treated with {\bf FullSimplify[ToGroebnerBasis}[\dots]].
\label{simpl}
}
: Reduction of ${}_{3}F_{2}$.
\\
\vspace{0.5cm}
{\bf ToGroebnerBasis}[ \{\{1+a$_1$,2+a$_2$, a$_3$\},\{1+b$_1$, b$_2$+2\},x\} ],
\\
\vspace{0.5cm}
{\small IntegerPart=\{1,2,0,1,2\} \,\,\,\,\,   changeVector=\{-1,-2,0,0,-1\}} \; ,
\\
\vspace{0.5cm}
$
\Biggl\{
\left\{
-\frac{\left(a_2-a_3+1\right) \left(b_2+1\right)}{\left(a_2+1\right) \left(a_3-b_2-1\right)},
-\frac{\left(b_2+1\right)
       \left(x a_2^2-\left(a_3 x-x+b_1\right) a_2+x a_1 \left(a_2-a_3+1\right)+b_1 b_2\right)}
      {x a_1 a_2 \left(a_2+1\right) \left(a_3-b_2-1\right)},
          -\frac{\left(b_2+1\right) \left(-a_3 x+x+(x-1) a_2+b_2\right)}
      {x a_1 a_2 \left(a_2+1\right)  \left(a_3-b_2-1\right)}
\right\} \;,
$
\\
\vspace{0.5cm}
$
\left\{\left\{a_1,a_2,a_3\right\},\left\{b_1+1,b_2+1\right\},x\right\},1
\Biggr \} \\
$
corresponds to
\begin{eqnarray}
&&
{}_{3}F_{2}\left(\begin{array}{c|}
 a_{1} + 1,  a_2 + 2,  a_3 \\
 b_{1} + 1 , b_{2}+2   \end{array} ~x \right)
=
\Biggl[
-\frac{\left(a_2-a_3+1\right) \left(b_2+1\right)}{\left(a_2+1\right) \left(a_3-b_2-1\right)}
\nonumber \\ &&
-\frac{\left(b_2+1\right)
   \left(x a_2^2-\left(a_3 x-x+b_1\right) a_2+x a_1 \left(a_2-a_3+1\right)+b_1 b_2\right)}{x a_1 a_2 \left(a_2+1\right)
   \left(a_3-b_2-1\right)}\theta
\nonumber \\ &&
-\frac{\left(b_2+1\right) \left(-a_3 x+x+(x-1) a_2+b_2\right)}
{x a_1 a_2 \left(a_2+1\right)
   \left(a_3-b_2-1\right)} \theta^2
\Biggr]
{}_{3}F_{2}\left(\begin{array}{c|}
 a_{1},   a_2,  a_3 \\
 b_{1}+1, b_{2}+1   \end{array} ~x \right) \;.
\end{eqnarray}

\item
{\bf Example 2}: Reduction of ${}_{4}F_{3}.$
\\
\vspace{0.5cm}
{\bf ToGroebnerBasis} [ \{\{1+a$_1$,1+a$_2$, a$_3$,a$_4$\},\{1+b$_1$, b$_2$+1,b$_3$\},x\} ] \;,
\\
\vspace{0.5cm}
{\small IntegerPart=\{1,1,0,0,1,1,0\}  \,\,\,\,\,   changeVector=\{-1,-1,0,0,0,0,1\}},
\\
\vspace{0.5cm}
$
\Biggl\{
\left\{1,\frac{1}{a_2}+\frac{1}{b_3}+\frac{1}{a_1},
         \frac{a_1+a_2+b_3}{a_1 a_2 b_3},
         \frac{1}{a_1 a_2 b_3}\right\},
\left\{\left\{a_1,a_2,a_3,a_4\right\},\left\{b_1+1,b_2+1,b_3+1\right\},x\right\},1
\Biggr \}
$\\
corresponds to
\begin{eqnarray}
&&
{}_{4}F_{3}\left(\begin{array}{c|}
  1+a_{1},1+a_2,a_3,a_4 \\
 1+b_{1},1+b_2,b_3 \end{array} ~x \right)
=
\Biggl[
1
+
\left( \frac{1}{a_2} + \frac{1}{b_3} + \frac{1}{a_1} \right) \theta
\nonumber \\ &&
+ \frac{a_1+a_2+b_3}{a_1 a_2 b_3} \theta^2
+ \frac{1}{a_1 a_2 b_3}\theta^3
\Biggr]
{}_{4}F_{3}\left(\begin{array}{c|}
 a_1,a_2,a_3,a_4 \\
 b_1+1,b_2+1,b_3+1 \end{array} ~x \right) \;.
\end{eqnarray}

\end{enumerate}
In both of these examples, $\mbox{IntegerPart}=\{\cdots\}$ corresponds to the set of integer values
of parameters in the original hypergeometric function, e.g.
$\mbox{IntegerPart}=\{1,2,0,1,2\}$ in {\tt Example 1},
and $\mbox{changeVector}=\{\cdots\}$ corresponds to the values of the parameters to be changed, e.g.
$\mbox{changeVector}=\{-1,-2,0,0,-1\}$ in {\tt Example 1}.

Routine {\bf explicitForm}[\dots] converts the results of the reduction, Eq.~(\ref{basis}),
to Mathematica-standard expressions for generalized hypergeometric functions.
\\
\\
{\bf Example 3} : Reduction of ${}_{2}F_{1}.$
\\
\vspace{0.5cm}
\mbox{answer}={\bf ToGroebnerBasis} [\{1+$a_1$,1+$a_2$\},\{1+$b_1$\},x\} ],
\\
\vspace{0.5cm}
{\small IntegerPart=\{1,1,1\}  \,\,\,\,\,   changeVector=\{-1,-1,0\}},
\\
\vspace{0.5cm}
{\bf explicitForm}[\mbox{answer}]
\\
\vspace{0.5cm}
$
\frac{\,\mathrm{ HypergeometricPFQ}\left(\{a_1,a_2\},\{b_1+1\},x\right)}{1-x}
+\frac{a_1 a_2 x \left(a_1+a_2-b_1\right) \,
\mathrm{HypergeometricPFQ}\left(\{a_1+1,a_2+1\},\{b_1+2\},x\right)}{\left(b_1+1\right) \left(a_1 a_2-a_1 a_2
   x\right)}.
$\\
For non-exceptional values of the parameters, the differential reduction is performed with the help of
Eqs.~(\ref{universal:a}), (\ref{universal:b}), (\ref{diff:oper:1}), and (\ref{diff:oper:2}).
The higher powers of the operators $\theta^k$   are expressed
with the help of the differential equation for the hypergeometric function, Eq.~(\ref{diffpFq}).
Also, the following relation is used in some cases:
\begin{equation}
{}_{p}F_{q}\left(\begin{array}{c|}
\{a_i  +  m\}_p \\
\{b_k  +  m\}_q  \end{array} ~z \right)
=
\frac{\prod_{k=1}^q\{ (b_k)_m \} }{\prod_{j=1}^p \{(a_j)_m \}}
\left(
\frac{d}{dz}
\right)^m
{}_{p}F_{q}\left(\begin{array}{c|}
\{a_i \}_p\\
\{b_k \}_q \end{array} ~z \right) \;.
\end{equation}

\subsection{Exceptional values of parameters}
When some of the upper parameters of the initial hypergeometric function are integer, then
the higher powers of the differential operator can be excluded with the  help of Eq.~(\ref{relation}).
In this case,
the input of {\bf ToGroebnerBasis}[\dots] does not change,
and the output of {\bf ToGroebnerBasis}[\dots] has the following structure
(in the case when only one upper parameter is integer):
\begin{gather}
\{\{\{Q_1,\cdots,Q_{p}\},\{\{1\!+\!a_1,\ldots,1\!+\!a_{p}\},\{2\!+\!b_1,\cdots,2\!+\!b_p\},x\},\mbox{factor}1\},
   \{Q_{p+1} , \{\}, \mbox{factor}2 \}\},
\label{output:2}
\end{gather}
where
\begin{itemize}
\item
$\{\{1+a_1,\cdots,1+a_{p+1}\},\{2+b_1,\ldots,2+b_p\},x\}$ are the set of parameters and argument of the resulting hypergeometric function;
\item
{\bf if an upper parameter is integer, the appropriate $a_j$ is equal to zero};
\item
$\{Q_1,\cdots,Q_{p}\}$ are the rational functions whose numbering corresponds to the power of $\theta^{i}$, $i=0,\ldots,p-1$;
\item
$\mbox{factor}$ is an overall factor;
\item
$Q_{p+1}$ is the resulting polynomial entering the r.h.s. of Eq.~(\ref{relation}).
\end{itemize}
This has the explicit form:
\begin{eqnarray}
&&
{}_{p+1}F_{p}\left(\begin{array}{c|}
 M_1, \cdots, M_r, a_{r+1}+M_{r+1}, \cdots,  a_{p+1} \!+\! M_{p+1} \\
 b_{1} \!+\! K_1, \cdots,  b_{p} \!+\! K_p   \end{array} ~x \right)
\nonumber \\ &&
=
\mbox{factor}1
\times
\biggl( Q_1 \!+\! Q_2\theta \!+\! \cdots \!+\! Q_{p} \theta^{p-1} \biggr)
{}_{p+1}F_{p}\left(\begin{array}{c|}
 1, \cdots, 1_r, 1\!+\!a_{r+1}, \cdots,  1\!+\!a_{p+1} \\
 2\!+\!b_{1} , \cdots, 2\!+\!b_{p}   \end{array} ~x \right)
\nonumber \\ &&
+ \mbox{factor}2 \times Q_{p+1}
\;,
\label{basis:2}
\end{eqnarray}
where $\vec{M}$ is a  set of integers.
\\
\\
{\bf Example 4} : Reduction of ${}_{3}F_{2}$ with an integer parameter.
\\
\vspace{0.5cm}
{\bf ToGroebnerBasis} [ \{\{3,1+a$_2$,1+a$_3$\},\{2+b$_1$,2+b$_2$\},x\} ],
\\
\vspace{0.5cm}
{\small IntegerPart=\{3,1,1,2,2\}  \,\,\,\,\,   changeVector=\{-2,0,0,0,0\}},
\\
\vspace{0.5cm}
$
\biggl\{\biggl\{\left\{\frac{\left(b_1+1\right) \left(b_2+1\right)
-x \left(a_2+1\right) \left(a_3+1\right)}{2 (x-1)}+1,\frac{1}{2} \left(\frac{-x
   \left(a_2+a_3+2\right)+b_1+b_2+2}{x-1}+3\right)\right\},
$
\\
\vspace{0.5cm}
$
 \left\{\left\{1,a_2+1,a_3+1\right\},\left\{b_1+2,b_2+2\right\},x\right\},  1\biggr\},
\{ -\frac{\left(b_1+1\right) \left(b_2+1\right)}{2 (x-1)}  ,\{\},  1\}\biggr\} \;.\\
$
This has the explicit form:
\begin{eqnarray}
&&
{}_{3}F_{2}\left(\begin{array}{c|}
3, a_2 \!+\!1, a_3 \!+\! 1 \\
b_1 \!+\! 2,b_2 \!+\! 2 \end{array} ~x \right)
=
\Biggl[
\frac{\left( b_1 \!+\!1 \right) \left(b_2 \!+\! 1 \right) \!-\! x \left( a_2\!+\!1 \right) \left(a_3 \!+\!1 \right)}{2 (x-1)}+1
\nonumber \\ &&
 + \frac{1}{2} \left(\frac{-x
   \left(a_2+a_3+2\right)+b_1+b_2+2}{x-1}+3\right)\theta
\Biggr]
{}_{3}F_{2}\left(\begin{array}{c|}
1,a_2 +1,a_3 + 1 \\
b_1 + 2,b_2 + 2 \end{array} ~x \right)
\nonumber \\ &&
- \frac{\left(b_1+1\right) \left(b_2+1\right)}{2 (x-1)} \;.
\nonumber
\end{eqnarray}

Other useful relations which are implemented in the function
{\bf ToGroebnerBasis}[\dots] include the relations derived in Ref.~\cite{karlsson}:
\begin{eqnarray}
&&
{}_{p}F_q\left(\begin{array}{c|}
b_1+m_1, \cdots, b_n+m_n, a_{n+1}, \cdots , a_p \\
b_1, \cdots, b_n, b_{n+1}, \cdots , b_q \end{array} ~z \right)
\nonumber \\ &&
=
\sum_{j_1=0}^{m_1}
\cdots
\sum_{j_n=0}^{m_n}
A(j_1, \cdots j_n) z^{J_n}
{}_{p-n}F_{q-n}\left(\begin{array}{c|}
a_{n+1}+J_n, \cdots a_p+J_n \\
b_{n+1}+J_n, \cdots , b_q+J_n \end{array} ~z \right) \; ,
\label{criterion1}
\end{eqnarray}
where $m_j$ are positive integers, $J_n = j_1 + \cdots +j_n$, and
\begin{eqnarray}
A(j_1, \cdots j_n)
&=&
\left( m_1 \atop j_1 \right)
\cdots
\left( m_n \atop j_n \right)
\frac{(b_2 \!+\! m_2)_{J_1} (b_3 \!+\! m_3)_{J_2} \cdots (b_n \!+\! m_n)_{J_{n \!-\! 1}} (a_{n \!+\! 1})_{J_n} \cdots (a_{p})_{J_n} }
     {(b_1)_{J_1} (b_2)_{J_2} \cdots (b_n)_{J_n} (b_{n \!+\! 1})_{J_n} \cdots (b_{q})_{J_n} } \;.
\nonumber \\
\end{eqnarray}
\\
\\
{\bf Example 5} : Reduction of ${}_{3}F_{2}$ with an integer difference of values of parameters.
\\
\vspace{0.5cm}
{\bf ToGroebnerBasis} [\{\{3+b$_1$,1+a$_2$,1+a$_3$\},\{2+b$_1$,2+b$_2$\},x\}],
\\
\vspace{0.5cm}
{\small IntegerPart=\{3,1,1,2,2\}  \,\,\,\,\,   changeVector=\{-1,-1,-1\}},
\\
\vspace{0.5cm}
$
\Biggl\{ \left\{-\frac{b_2+1}{(x-1) \left(b_1+2\right)},
  -\frac{\left(a_2 x+a_3 x-b_1 x-x+b_1-b_2+1\right)
   \left(b_2+1\right)}{(x-1) x a_2 a_3 \left(b_1+2\right)}\right\},
   \left\{\left\{a_2,a_3\right\},\left\{b_2+1\right\},x\right\},  1 \Biggr\}.
$
\\
This has the explicit form:
\begin{eqnarray}
&&
{}_{3}F_{2}\left(\begin{array}{c|}
 3+b_{1},1+a_2,1+a_3 \\
 2+b_{1},2+b_2 \end{array} ~x \right)
\\ &&
=
\Biggl[
-\frac{b_2 \!+\! 1}{(x-1) \left(b_1 \!+\! 2\right)}
-\frac{\left(a_2 x \!+\! a_3 x \!-\! b_1 x \!-\! x \!+\! b_1 \!-\! b_2 \!+\! 1\right)
       \left(b_2 \!+\! 1\right)}{(x\!-\!1) x a_2 a_3 \left(b_1 \!+\! 2\right)}\theta
\Biggr]
{}_{2}F_{1}\left(\begin{array}{c|}
 a_2,a_3 \\
 b_2 \!+\! 1 \end{array} ~x \right) \;.
\nonumber
\end{eqnarray}

The following relations are implemented in Mathematica 7.0:
\begin{eqnarray}
&&
\theta^k
{}_{p+1}F_{p}\left(\begin{array}{c|}
A,         \vec{a} \\
1  +  A, \vec{b}  \end{array} ~z \right)
=
(-A)^k
{}_{p+1}F_{p}\left(\begin{array}{c|}
A,         \vec{a} \\
1  +  A, \vec{b}  \end{array} ~z \right)
-
\sum_{j=0}^{k-1}
(-A)^{k-j} \theta^j
{}_{p}F_{p-1}\left(\begin{array}{c|}
\vec{a} \\
\vec{b}  \end{array} ~z \right)
\;,
\nonumber \\ &&
\left(
\frac{\theta}{A}
\right)^q
{}_{p+1}F_{p}\left(\begin{array}{c|}
\{A \}_r,        \vec{a} \\
\{1  +  A \}_r, \vec{b}  \end{array} ~z \right)
=
\sum_{j=0}^q (-1)^{(j+q)}
\left( q \atop j\right)
{}_{p+1-j}F_{p-j}\left(\begin{array}{c|}
\{A \}_{r-j},        \vec{a} \\
\{1  +  A \}_{r-j}, \vec{b}  \end{array} ~z \right) \;,
\nonumber \\ &&
{}_{p}F_q\left(\begin{array}{c|}
1, \{a_i \}_{p-1} \\
2, \{b_k \}_{q-1}
\end{array} ~z \right)
=
\frac{1}{z}
\frac{\prod_{l=1}^{q-1} (b_l - 1)}
     {\prod_{j=1}^{p-1} (a_j - 1)}
\left[
{}_{p-1}F_{q-1}\left(\begin{array}{c|}
\{a_i-1 \}_{p-1} \\
\{b_k-1 \}_{q-1}
\end{array} ~z \right)
- 1
\right] \;,
\end{eqnarray}
where $q \leq r$
and
$a_j,b_k \neq 1.$

\section{AppellF1F4 - Mathematica-based program for differential reduction of Appell's functions $F_1,F_2,F_3,F_4$}
\label{AppellF1F4}

In this section, we will present the MATHEMATICA-based\footnote{It was tested for MATHEMATICA $7.0$.}
program {\bf AppellF1F4}  for the differential reduction of the Appell hypergeometric functions
$F_1$, $F_2$, $F_3$, and $F_4$.
The program is available from Ref.~\cite{bytev:hyper}.
The current version, $1.0$, only deals  with non-exceptional values of parameters.

The program may be loaded in the standard way:
$$
<< \mathrm{"AppellF1F4.m"}
$$
The package includes the following basic routines:
\begin{eqnarray}
&&
{\bf F1IndexChange}[\mbox{changingVector}, \mbox{parameterVector}],
\label{input:f1}
\\ &&
{\bf F2IndexChange}[\mbox{changingVector}, \mbox{parameterVector}],
\label{input:f2}
\\ &&
{\bf F3IndexChange}[\mbox{changingVector}, \mbox{parameterVector}],
\label{input:f3}
\\ &&
{\bf F4IndexChange}[\mbox{changingVector}, \mbox{parameterVector}],
\label{input:f4}
\\ &&
{\bf explicitFormF1}[\dots] \;.
\end{eqnarray}

The $"{\rm changingVector}"$  in Eqs.~(\ref{input:f1})--(\ref{input:f4})
is the set of integers  at which we wish to change the values of parameters of the  Appell functions
(the vector $\vec m$ in Eqs.~(\ref{reductionf1}) and (\ref{reductionf234}) ).
The set of parameters  of the Appell function are defined in the list $"{\rm parameterVector}"$
(corresponding to the vector $\vec{A}+\vec{m}$ and the arguments $x,y$ on the l.h.s. of Eqs.~(\ref{reductionf1}) 
and (\ref{reductionf234}) ).
We wish to point out that the enumeration of the parameters and arguments in the list 
$"{\rm parameterVector}"$
 corresponds one-to-one to Eqs.~(\ref{definition:f1}), (\ref{definition:f2}), (\ref{definition:f3}), 
and (\ref{definition:f4}).

The output of ${\bf F1IndexChange}[]$ and ${\bf F\{234\}IndexChange}[]$
has a somewhat different structure.

\subsection{Appell function $F_1$}
The  structure of the output of ${\bf F1IndexChange}[]$ is the following:
\begin{eqnarray}
\{\{\mbox{A,B,C} \},\{ {\rm parameterVectorNew} \},\{ {\rm AppellF1} \}\},
\end{eqnarray}
where
\begin{enumerate}
\item
${\rm parameterVectorNew} $ is the set of new parameters of the Appell function $F_1$ ;
\item
${\mbox A,B,C}$ are the rational functions corresponding to
the ratios $P_0/R, P_1/R$, and $P_2/R$ of the functions entering Eq.~(\ref{reductionf1}).
\end{enumerate}
{\bf Example 6}
\footnote{
See footnote \ref{simpl}.
}:
Reduction of $F_1$.
\\
\\
{\bf F1IndexChange[}\{$1$,$-1$,$0$,$0$\},\,\{$a$,$b_1$,$b_2$,$c$,$z_1$,$z_2$\}{\bf ]},
\\
\\
$\left\{\left\{\frac{a (-\text{z1})+a+\text{b1} \text{z$_1$}+\text{b$_2$} \text{z$_2$}-c-\text{z$_1$}+1}{a-c+1},
-\frac{(\text{z$_1$}-1) (a-\text{b$_1$}+1)}{(\text{b$_1$}-1)
   (a-c+1)},\frac{\text{z$_2$}-1}{a-c+1}\right\},\{a+1,\text{b$_1$}-1,\text{b$_2$},c,\text{z$_1$},\text{z$_2$}\},
\text{AppellF1}\right\}
$.
\\
\\
This has the explicit form:
\begin{eqnarray}
&&
F_1(a,b_1,b_2,c;z_1,z_2)
=
\nonumber \\ &&
\Biggl[
  \frac{- a z_1 + a + b_1 z_1 + b_2 z_2 - c - z_1 + 1 }{a-c+1}
- \frac{(z_1-1) (a-b_1+1)}{(b_1-1) (a-c+1)} \theta_1
+ \frac{z_2-1}{a-c+1}\theta_2
\Biggr]
\nonumber \\ &&
\times
F_1(a+1,b_1-1,b_2,c; z_1,z_2).
\end{eqnarray}

Routine {\bf explicitFormF1}[$\cdots$] converts the result of the reduction to a Mathematica standard expression
for the Appell function $F_1$.
\\
\\
{\bf Example 7} : Reduction of $F_{1}$.
\\
\mbox{result}  = {\bf F1IndexChange[}\{$1$,$-1$,$0$,$0$\},\,\{$a$,$b_1$,$b_2$,$c$,z$_1$,z$_2$\}{\bf ]},
\\
\\
{\bf explicitFormF1}[\mbox{result}]
\\
\\
$
-\frac{(a+1) \left(z_1-1\right) z_1 \left(a-b_1+1\right)}{c (a-c+1)} \mathrm{AppellF1}\left(a+2;b_1,b_2;c+1;z_1,z_2\right)
$
\\
$
+\frac{(a+1) b_2 \left(z_2-1\right) z_2}{c (a-c+1)}
  \mathrm{AppellF1}\left(a+2;b_1-1,b_2+1;c+1;z_1,z_2\right)
$
\\
$
+\frac{\left(a \left(-z_1\right)+a+b_1 z_1+b_2 z_2-c-z_1+1\right)}{a-c+1}\mathrm{AppellF1}\left(a+1;b_1-1,b_2;c;z_1,z_2\right) \;.
$

\subsection{Appell functions $F_2$, $F_3$, $F_4$}
The outputs of
${\bf F2IndexChange}[]$,
${\bf F3IndexChange}[]$,
and
${\bf F4IndexChange}[]$
are similar and have the following structure:
\begin{eqnarray}
\{\{\mbox{A,B,C,D} \},\{ {\rm parameterVectorNew} \},\{ {\rm NameOfFunction} \},
\end{eqnarray}
where
\begin{enumerate}
\item
${\rm NameOfFunction} $ is the name of the Appell functions to be reduced:
$\mbox{AppellF2},
\mbox{AppellF3},
$
or
$
\mbox{AppellF4}
$ ;
\item
${\rm parameterVectorNew} $ is the set of new parameters of the Appell function;
\item
${\mbox A,B,C,D}$ are the rational functions corresponding to
the ratios  $Q_0/S, Q_1/S, Q_2/S$, and $Q_3/S$ of the functions entering Eq.~(\ref{reductionf234}).
\end{enumerate}
{\bf Example 8} : Reduction of $F_{2}$.
\\
\\
{\bf F2IndexChange[}\{$0$,$0$,$1$,$1$,$0$\},\,\{a,b$_1$,b$_2$,c$_1$,c$_2$,z$_1$,z$_2$\}{\bf ]},
\\
\\
$\left\{\left\{\frac{a z_2 \left(c_1 \left(z_1-1\right)-b_1 z_1\right)}{c_1
  \left(z_1-1\right) \left(b_2-c_2+1\right)}+1,\frac{1-\frac{z_2 \left(a+z_1
   \left(b_1-c_1\right)\right)}{\left(z_1-1\right)
    \left(b_2-c_2+1\right)}}{c_1},\frac{c_1 \left(z_1-1\right) \left(z_2-1\right)-b_1 z_1 z_2}{c_1 \left(z_1-1\right)
   \left(b_2-c_2+1\right)},-\frac{z_1+z_2-1}{c_1 \left(z_1-1\right) \left(b_2-c_2+1\right)}\right\},\right.
$
\\
$
\left.
\left\{a,b_1,b_2+1,c_1+1,c_2,z_1,z_2\right\},\text{AppellF2}\right\}
$.
\\
\\
This has the explicit form:
\begin{eqnarray}
&&
F_2(a,b_1,b_2,c_1,c_2,z_1,z_2)
=
\nonumber \\ &&
\Biggl[
\frac{a z_2 \left(c_1 \left(z_1-1\right)-b_1 z_1\right)}{c_1 \left(z_1-1\right) \left(b_2-c_2+1\right)}+1
+\frac{1-\frac{z_2 \left(a+z_1
   \left(b_1-c_1\right)\right)}{\left(z_1-1\right) \left(b_2-c_2+1\right)}}{c_1}\theta_1
+\frac{c_1 \left(z_1-1\right) \left(z_2-1\right)-b_1 z_1 z_2}{c_1 \left(z_1-1\right)
   \left(b_2-c_2+1\right)}\theta_2
\nonumber \\ &&
-\frac{z_1+z_2-1}{c_1 \left(z_1-1\right) \left(b_2-c_2+1\right)}\theta_1\theta_2
\Biggr]
F_2(a,b_1,b_2+1,c_1+1,c_2;z_1,z_2) \;.
\end{eqnarray}

The Appell functions $F_2$, $F_3$, and $F_4$ are not built in the
current version of Mathematica (version 7.0), so that the
{\bf AppellF1F4} package does not include a function similar to {\bf explicitFormF1[]}.

\section{Conclusion }
\label{conclusion}

The differential-reduction algorithm \cite{theorem} allows one to  compare
Horn-type hypergeometric functions with parameters whose values differ
by integers.
In this paper, we presented the Mathematica-based package
{\bf HYPERDIRE} for the differential reduction of the
generalized hypergeometric functions $_{p+1}F_p$ and the Appell functions
$F_1,F_2,F_3,$ and $F_4$ to  sets of basis functions, defined
by Eqs.
(\ref{decomposition}),
(\ref{reductionf1}),
(\ref{reductionf234}),
respectively.
These functions are closely related to a large class of Feynman diagrams \cite{FD}.
In contrast to existing packages, our package performs the reduction of hypergeometric functions
before the $\ep$ expansion and works with  arbitrary values of parameters.
This package could be easily extended to other Horn-type hypergeometric functions by adding
new modules.

As an illustration of our approach,
we considered a few examples \cite{reduction:our:1,reduction:our:2}
with arbitrary powers of propagators and space-time dimension \cite{dimreg}.


\vspace{5mm}
\noindent
{\bf Acknowledgments.} \\
We are grateful to
A.~Davydychev,
A.~Grozin,
A.~Isaev,
A.~Kotikov,
G.~Somogyi,
V.~Spiridonov,
O.~Tarasov,
O.~Veretin,
B.F.L~Ward 
and 
S.~Yost for useful discussions,
and to 
G.~Sandukovskaya
for carefully reading the manuscript.
The work of V.V.B. was supported in
part by the Russian Foundation for Basic Research RFFI through Grant
No.~12-02-31703 and by the Heisenberg-Landau Program.
This work was supported in part by the German Federal Ministry for Education
and Research BMBF through Grant No.\ 05H12GUE and by the German Research
Foundation DFG through the Collaborative Research Centre No.~676
{\it Particles, Strings and the Early Universe---The Structure of Matter and
Space-Time}.


\end{document}